\documentclass[fleqn,usenatbib,useAMS]{mnras}
\usepackage{graphicx}	
\usepackage{amsmath}	
\usepackage{amssymb}	
\usepackage{multicol}   
\usepackage{bm}		
\usepackage{pdflscape}	
\usepackage{soul}
\usepackage{appendix}
\usepackage{ulem}
\usepackage{textcomp}
\usepackage{tcolorbox}
\usepackage{caption}

\usepackage{multibib}
\newcommand{\beq}{\begin{equation}}
\newcommand{\eeq}{\end{equation}}
\newcommand{\bse}{\begin{subequations}}
\newcommand{\ese}{\end{subequations}}
\newcommand{\bary}{\begin{eqnarray}}
\newcommand{\eary}{\end{eqnarray}}
\newcommand{\bwt}{\begin{widetext}}
\newcommand{\ewt}{\end{widetext}}

\usepackage[T1]{fontenc}
\usepackage{ae,aecompl}

\title{Constraining the redshift of PG 1553+113 using the photohadronic model}

\author[Sarira Sahu et al.]
{
Sarira Sahu$^{1}$
\thanks{Contact e-mail: \href{mailto:sarira@nucleares.unam.mx}{sarira@nucleares.unam.mx}},
R. de J.~Pacheco-Aké$^{2}$
\thanks{Contact e-mail: \href{mailto:rodrigo.pacheco@cinvestav.mx }{rodrigo.pacheco@cinvestav.mx}},
G.~Sánchez-Colón$^{2}$
\thanks{Contact e-mail: \href{mailto: gabriel.sanchez@cinvestav.mx}{gabriel.sanchez@cinvestav.mx}},
D. I. Páez-Sánchez$^{1}$
\thanks{Contact e-mail: \href{mailto:diana.paez@correo.nucleares.unam.mx}{diana.paez@correo.nucleares.unam.mx}},
\newauthor
A.~U.~Puga Oliveros$^{1}$
\thanks{Contact e-mail: \href{mailto: angel.puga@correo.nucleares.unam.mx}{angel.puga@correo.nucleares.unam.mx}}
Subhash Rajpoot$^{3}$
\thanks{Contact e-mail: \href{mailto: Subhash.Rajpoot@csulb.edu}{Subhash.Rajpoot@csulb.edu}},
\\
$^{1}$Instituto de Ciencias Nucleares, Universidad Nacional Aut\'onoma de M\'exico,
Circuito Exterior S/N, C.U., A.P. 70-543,\\ CDMX 04510, México.\\
$^{2}$Departamento de Física Aplicada, Centro de Investigación y de Estudios Avanzados del IPN, Unidad Mérida.
A.P. 73, Cordemex,\\ Mérida, Yucatán 97310, México.\\
$^{3}$Department of Physics and Astronomy, California State University, 1250 Bellflower Boulevard, Long Beach, CA 90840, USA.
}

\date{}

\pubyear{2024}

\begin{document}
\label{firstpage}
\pagerange{\pageref{firstpage}--\pageref{lastpage}}
\maketitle


\begin{abstract}

Between 2005 and 2015, the BL Lacertae object PG 1553+113 exhibited multiple very high-energy (VHE; \(>100\) GeV) gamma-ray flares, which were detected by the Cherenkov telescopes, High Energy Stereoscopic System (HESS), Major Atmospheric Gamma Imaging Cherenkov (MAGIC), and the Very Energetic Radiation Imaging Telescope Array System (VERITAS). Despite the uncertainty surrounding its redshift (\(z\)), various studies have sought to estimate this value. In this study, seventeen independently observed VHE gamma-ray spectra of PG 1553+113 are analysed using four distinct extragalactic background light (EBL) models alongside the photohadronic framework. A global \(\chi^2\) fit is applied to all observational data to determine the best-fitting redshift for each EBL model. Additionally, confidence level (CL) intervals for the redshift are calculated across all EBL models. The findings demonstrate that the photohadronic framework effectively describes all observed spectra. Among the EBL models, the lowest total \(\chi^2\) value of 58.82 (with 58 degrees of freedom) was obtained using the model from Saldana-Lopez et al. (2021), while the highest value of 75.67 was associated with the model from Domínguez et al. (2011). The 95 per cent CL intervals for the statistical error of the redshift of PG 1553+113 are \(0.500 < z < 0.537\) for the Saldana-Lopez model and \(0.491 < z < 0.527\) for the Domínguez  model. These results highlight the consistency of the photohadronic approach in interpreting the observed VHE gamma-ray spectra.

\end{abstract}

\begin{keywords}
{BL Lacertae objects: individual: PG 1553+113, galaxies: distances and redshifts, methods: statistical}
\end{keywords}
\color{black}

\section{INTRODUCTION}\label{Intro}

BL Lacertae (BL Lac) blazars represent a unique category of active galactic nuclei (AGN), distinguished by their non-thermal emission spectra, which arise from a relativistic jet oriented nearly parallel to the observer's line of sight~\citep{Urry_1995, VERITAS:2010vjk}. These objects display rapid variability across the entire electromagnetic spectrum. Their spectral energy distributions (SEDs) are marked by a characteristic two-peak structure~\citep{Abdo:2009iq}. The low-energy peak is attributed to the synchrotron radiation emitted by the low-energy electrons as they traverse the jet's magnetic field. The high-energy peak, on the other hand, can be either due to the inverse Compton scattering of high-energy electrons with their self-produced synchrotron photons—referred to as synchrotron self-Compton (SSC) scattering~\citep{1992ApJ...397L...5M,1993ApJ...416..458D,1994ApJ...421..153S,Blazejowski:2000ck,Murase_2012,Gao:2012sq}—or the scattering of external photons originating from the accretion disk, broad-line regions, or the dusty torus surrounding the AGN~\citep{1993ApJ...416..458D,1994ApJ...421..153S,Blazejowski:2000ck}.

In BL Lac objects, the non-thermal radiation significantly overshadows the stellar light emitted by their host galaxies, making it challenging to determine accurately the redshifts. This limitation creates uncertainties in studying the cosmic evolution, deciphering the source's characteristics, and analysing its intrinsic VHE spectrum. Given that VHE photons are absorbed by the EBL, obtaining precise redshift measurements is essential for understanding the EBL's influence on VHE photons.

PG 1553+113, a well-known blazar, was first identified in a photographic survey~\citep{1986ApJS...61..305G}. Its spectral properties have been thoroughly examined, leading to its classification as a high-frequency synchrotron peaked BL Lac (HBL)~\citep{2019A&A...632A..77C}. Over the years, multiple VHE observations have been carried out using different instruments. The HESS telescope array first detected this blazar in gamma-rays in 2005~\citep{2006AA448L19A}, and it was later monitored by MAGIC and VERITAS in various campaigns between 2005 and 2015, especially during periods of heightened activity. PG 1553+113 is a prime target for high-energy astrophysics due to its extreme variability and emission mechanisms. However, its redshift remains unconfirmed via spectroscopy, making indirect estimation methods critical for interpreting its emission properties and cosmological impact.

The VHE gamma-ray flux from a blazar ($F_\gamma$), as measured on Earth, can be expressed as~\citep{Hauser:2001xs}:
\begin{equation}
F_{\gamma}(E_{\gamma}) = F_{\rm int}(E_\gamma) \, e^{-\tau_{\gamma\gamma}(E_{\gamma}, z)}, 
\label{eq:flux}
\end{equation}
where \( E_\gamma \) denotes the energy of the observed VHE photon, \( F_{\gamma} \) is the observed flux, and \( F_{\rm int} \) represents the intrinsic flux. The optical depth, \( \tau_{\gamma \gamma} \), which arises from the pair production process \( \gamma\gamma \rightarrow e^+e^- \), is a function of both \( E_\gamma \) and the redshift \( z \) of the source. The exponential term in Eq. (\ref{eq:flux}) signifies the reduction in VHE flux caused by \( e^+e^- \) pair production~\citep{1992ApJ...390L..49S,doi:10.1126/science.1227160,Padovani:2017zpf}. As a result, determining the redshift with precision is critical for deriving the intrinsic flux from the observed measurements. To analyse VHE gamma-ray spectra from sources at different redshifts, the Imaging Atmospheric Cherenkov Telescopes (IACTs) collaborations rely on well-established EBL models developed by~\cite{Franceschini:2008tp,2010ApJ...712..238F,Dominguez:2010bv,10.1111/j.1365-2966.2012.20841.x,2021MNRAS.507.5144S}.

PG 1553+113 lacks a direct spectroscopic redshift due to its featureless optical spectrum (a hallmark of BL Lacs) but has been inferred through various methods. Using data from HESS observations in 2005 and considering the absorption of VHE gamma-rays by the EBL, \cite{2006AA448L19A} derived an upper limit of $z < 0.74$. Later, \cite{2010ApJ720976D} established a lower limit of $0.4 < z$ by detecting intervening Ly\ $\alpha$ absorbers, while also setting an upper limit of $z < 0.58$ due to the absence of Ly\ $\beta$ absorbers at $z > 0.4$. Subsequent studies have supported these constraints. For instance, \cite{2015ApJ7997A} analysed the time-averaged VHE spectrum from VERITAS observations between 2010 and 2012, incorporating EBL evolution, and proposed a redshift upper limit of $z \leq 0.62$. More recently, \cite{2022MNRAS.509.4330D} used a Ly\ $\alpha$-forest-based approach to suggest a narrower redshift range of $0.408 < z < 0.436$. Our study aims to estimate the redshift of PG 1553+113 using the  photohadronic model, employing a statistical approach and a comprehensive dataset.

The photohadronic model~\citep{Sahu:2019lwj} has previously been applied successfully to interpret the VHE gamma-ray spectra of HBLs and extreme HBLs (EHBLs)~\citep{2020ApJ...901..132S,2021ApJ...914..120S,2022MNRAS.515.5235S}. In this work, the photohadronic model~\citep{Sahu:2019lwj} is combined with four established EBL models (\cite{Franceschini:2008tp,Dominguez:2010bv,10.1111/j.1365-2966.2012.20841.x,2021MNRAS.507.5144S}) to analyse seventeen VHE flaring events of PG 1553+113 observed between 2005 and 2015 by the HESS, VERITAS, and MAGIC collaborations. By performing a global $\chi^2$ fit to the experimental data from these VHE spectra, central values and CL intervals for the redshift of PG 1553+113 are determined for each EBL model. This method has been validated in prior analyses, such as the redshift constraint determination for the BL Lac object PKS 1424+240, which aligned with spectroscopic measurements~\citep{10.1093/mnras/stae1847}.

The structure of this paper is organized as follows. Section~\ref{sample} provides an overview of the seventeen independent observations of PG 1553+113 used in the analysis. Section~\ref{sec3} offers a concise review of the photohadronic model and its relevance to this study. Section~\ref{sec4} presents the analysis and results, while Section~\ref{sec5} concludes with a summary and discussion of the findings.

\section{VHE Observations of PG 1553+113}\label{sample}

This section provides a concise overview of the seventeen observations of blazar PG 1553+113 analysed in this study to constrain its redshift. The data were collected by three major Cherenkov telescopes: MAGIC~\citep{2004MmSAI..75..232M}, HESS~\citep{2004NewAR..48..331H}, and VERITAS~\citep{2006APh....25..391H}. Key details for each observation are summarized in Table~\ref{table1}. To simplify referencing, each observation is labelled according to the telescope and the year it was conducted. For instance, the HESS observation of 2005 is labeled as H05.

\subsection{H05, H06}

In reference~\cite{2008AA477481A}, a re-analysis of the 2005 HESS data (7.6 hours) for PG 1553+113~\citep{2006AA448L19A} is presented, incorporating an improved calibration of the detector's absolute energy scale. Additionally, results from 17.2 hours of HESS observations conducted in 2006 are reported.

The HESS system observed PG 1553+113 in the VHE gamma-ray regime during observational campaigns in 2005 and 2006. Over the two-year period, HESS accumulated 24.8 hours of live time, resulting in the detection of a VHE gamma-ray signal with a statistical significance of approximately 10 standard deviations. The emission displayed remarkable stability, with no significant flux variability or spectral changes observed across any timescale.
   
Complementary near-infrared spectroscopic observations targeting the $H+K$ band (1.45–2.45 $\mu$m) were conducted in March 2006 using the Spectrograph for INtegral Field Observations in the Near Infrared (SINFONI) on the European Southern Observatory’s Very Large Telescope (VLT) in Chile. Despite the exceptional sensitivity of the SINFONI integral field spectrometer, producing one of the highest-resolution near-infrared spectra of PG 1553+113 to date, the data revealed no detectable absorption or emission features. This absence of spectral lines precluded any direct constraints on the source’s redshift, leaving its cosmological distance uncertain.

A total of 30.3 hours of HESS observations were conducted, typically in segments of about 28 minutes (runs), during 2005 and 2006. The telescopes were offset by $\pm 0.5^{\circ}$ from the position of PG 1553+113 during these observations. After applying data-quality selection criteria, the useful exposure time amounted to 24.8 hours of live time, with 9 of the 66 runs excluded. The 2005 observation, referred to as H05, had a live time of 7.6 hours, while the 2006 observation, labelled H06, had a live time of 17.2 hours.

\subsection{H12}

The HESS telescopes observed significant flaring activity in the VHE gamma-ray emissions from the HBL PG 1553+113. Notably, the source's flux enhanced by a factor of about three on the nights of April 26 and 27, 2012, compared to previous archival measurements, with indications of variability within the same night. Interestingly, during this period ${\it Fermi}$ Large Area Telescope did not detect any signal. Using Bayesian statistical approach to the above data, the redshift was found to be $z = 0.49 \pm 0.04$~\citep{2015ApJ80265A}.

As discussed previously, the observations H05 and H06 showed no signs of variability and are collectively termed as the "pre-flare" data set. This pre-flare data set comprises 26.4 hours of high-quality live time observations. During the 2012 flaring period (referred here as H12), eight observational runs, each lasting approximately 28 minutes were conducted on April 26 and 27, totalling 3.5 hours of live time. All observations were performed in wobble mode, where the source is positioned at an offset of 0.5$^\circ$ from the centre of the instrument's field of view. This method resulted in acceptance-corrected live times of 24.8 hours for the pre-flare data and 3.2 hours for the flare data.

\subsection{H13}

The HESS phase II, with a fifth telescope CT5, observed the HBL PG 1553+113 in 2013 between May 29 and August 9, 2013 (MJD 56441–56513)~\citep{2017AA600A89H}, referred to as H13 in this work. These observations are divided into 28-minute segments called runs and conducted in wobble mode, where the camera's field of view is offset by $0.5^\circ$ or $0.7^\circ$ from the source position, either along the right-ascension or declination axis. For this analysis, only runs with the source positioned between $0.35^\circ$ and $1.2^\circ$ from the camera centre were included. During the commissioning phase, some runs employed non-standard wobble offsets to evaluate the instrument's performance. This approach ensures the source remains well within the field of view and facilitates accurate background subtraction. The analysis comprises 39 runs with 16.8 hours of live time.

\subsection{M0506}

The MAGIC telescope detected VHE gamma-ray emissions from the HBL PG 1553+113 during observations conducted in 2005 and 2006. The signal achieved an overall significance of 8.8 standard deviations over a total observation time of 18.8 hours. The spectrum did not exhibit significant daily flux variations and the flux levels in 2005 were notably higher than those recorded in 2006. The combined differential energy spectrum for PG 1553+113 from these two years is presented in reference~\cite{Albert_2007} and is referred to as M0506 in this work. Observations of PG 1553+113 with the MAGIC telescope took place in two main periods: 8.9 hours in April and May 2005, coinciding with HESS observations~\citep{2008AA477481A}, and 19 hours from January to April 2006.

To ensure data quality, observations affected by poor weather conditions or hardware issues were excluded from the analysis. Additionally, only data collected at small zenith angles (< 30°), which correspond to a lower energy threshold suitable for studying steep energy spectra, were included. Although measurements extended up to zenith angles of 53°, the final analysis retained 7.0 hours of high-quality data from 2005 and 11.8 hours from 2006 after applying these selection criteria.

\subsection{M06}

The MAGIC telescope observed PG 1553+113 during a multiwavelength campaign from July 14 to July 27, 2006, accumulating 9.5 hours of data at zenith angles ranging between $18^\circ$ and $35^\circ$~\citep{2009AA493467A}. These observations are labelled as M06 in this study. The data were collected in wobble mode, with the source positioned at an offset of $\pm0.4^\circ$ from the camera's centre, allowing simultaneous measurement of both the source and the background. Approximately one hour of data was discarded due to technical issues.

The quality of the data set was impacted by calima, a phenomenon involving Saharan sand-dust in the atmosphere, which caused nightly atmospheric absorption levels to vary between 5 and 40 per cent. To address the attenuation of Cherenkov light, correction factors were computed and applied to the affected data. After these adjustments, a signal with a significance of 5.0 standard deviations was detected in 8.5 hours of observation.

\subsection{M07, M08, M09}

Following its initial detection, PG 1553+113 became a regular target for observations with the MAGIC telescope. As reported in reference~\cite{2012ApJ74846A}, new data collected in 2007, 2008, and 2009 were analysed, and here we refer to these observations as M07, M08, and M09, respectively. Since the autumn of 2009, MAGIC has operated as a stereo system consisting of two IACTs located on La Palma in the Canary Islands, Spain. However, the data discussed here were acquired before this upgrade, using only the single MAGIC I telescope.

As mentioned previously, PG 1553+113 was observed by MAGIC for approximately 19 hours in 2005 and 2006~\citep{Albert_2007} and was also the focus of a multiwavelength campaign in July 2006 involving optical, X-ray, and TeV gamma-ray ~\citep{2009AA493467A}. Reference~\cite{2012ApJ74846A} presents the results of follow-up observations conducted for 14 hours in March–April 2007, nearly 26 hours in March–May 2008 (some of which were simultaneous with observations by other instruments~\citep{2010AA515A76A}), and about 24 hours in March–July 2009, with some of the latter taken under moderate moonlight conditions. Unfortunately, the 2008 and 2009 observations were significantly impacted by adverse weather, including calima (Saharan sand dust in the atmosphere), which reduced the usable data and increased the energy threshold.

Rigorous quality cuts based on event rates after night sky background suppression were applied to the data. After these cuts, 28.7 hours of high-quality data remained, comprising 11.5 hours from 2007, 8.7 hours from 2008, and 8.5 hours from 2009. Combining the results from these three years, the 28.7 hours of observations yielded a signal with an overall significance of 8.8 standard deviations. The individual significances were 5.8 standard deviations for 2007, 8.1 for 2008, and 4.2 for 2009.

\subsection{M08II}

A comprehensive simultaneous multi-frequency observational campaign was conducted between March and April 2008, involving data collection across optical, X-ray, high-energy (HE) gamma-ray, and VHE gamma-ray bands~\citep{2010AA515A76A}. The average VHE differential spectrum of PG 1553+113 derived from these observations is reported and labelled as M08II in this work.

The MAGIC observations for this campaign took place on March 16–18 and April 13, 28–30, 2008, with zenith angles ranging from 18 to 36 degrees. Observations were conducted in wobble mode, where the source was positioned at a 0.4-degree offset from the camera centre, alternating directions every 20 minutes. After applying standard quality cuts and filtering based on trigger rates, a total of 7.18 hours of effective observation time was retained for analysis.

To reconstruct the gamma-ray spectrum, looser selection criteria were employed to ensure that over 90 per cent of simulated gamma photons were preserved. The impact of varying cut efficiencies, ranging from 50 to 95 per cent across the energy spectrum, was examined to assess systematic effects on the spectral shape. Corrections were applied to data affected by calima.

Analysis of the MAGIC data revealed an excess of 415 gamma-like events compared to 1835 normalized background events, resulting in a detection significance of 8.0 standard deviations.

\subsection{M12}

In early 2012, PG 1553+113 entered a high state, and by April of that year, it reached its highest recorded VHE flux level to date. A comprehensive multiwavelength observation campaign was conducted from February to June 2012, aimed at characterizing the SED and studying the variability of the source across different frequency bands with observations spanned from VHE gamma-rays to radio wavelengths. 

Reference~\cite{2015MNRAS.450.4399A} details the analysis of flux variability in the VHE, HE, and X-ray bands. MAGIC observed PG 1553+113 from February 26 (MJD 55983) to April 26 (MJD 56043), 2012. After applying quality cuts, the data set comprised 18.3 hours of observations with zenith angles ranging from 17 to 34 degrees. Observations were conducted in wobble mode, with the source positioned 0.4 degrees from the camera centre. The standard MAGIC analysis chain was used to process the data, yielding an energy threshold of approximately 70 GeV. The source was detected with a high statistical significance (> 70 standard deviations) during the February–April 2012 period, with the emission consistent with a point-like source at the location of PG 1553+113. Based on the flux levels, the data were divided into two subsets: MJD 55983–MJD 56016 (high state) and MJD 56037–MJD 56043 (flare state). During the flare, the VHE flux nearly doubled compared to the high state. The analysis of the VHE spectrum focuses on the April flare state, which is referred to as M12 in this work.

\subsection{M12II, M12III, M13, M14, M15}

In reference~\cite{10.1093/mnras/stz943}, a total of 32 MAGIC spectra were utilized to derive constraints on the EBL through a joint likelihood analysis of 12 blazars observed during extensive campaigns, amounting to over 300 hours of exposure. All data included in this study were collected during dark nights under optimal weather conditions.

Among these, five data sets correspond to the BL Lac object PG 1553+113, covering observations from 2012 to 2015 with a combined observation time of 66.36 hours. These data sets, as reported in reference~\cite{10.1093/mnras/stz943}, are labelled as PG1553+113\_0202 (referred to here as M12II), PG1553+113\_0203 (M12III), PG1553+113\_0302 (M13), PG1553+113\_0303 (M14), and PG1553+113\_0306 (M15).

\subsection{V1012}

The time-averaged VERITAS spectrum for the BL Lac object PG 1553+113, covering observations from 2010 to 2012 (referred to in this work as V1012), is analyzed in reference~\cite{2015ApJ7997A}. 

VERITAS, an IACTs array, can detect PG 1553+113 at energies above 100 GeV with a significance of 5 standard deviations in approximately 43 minutes of observation, based on its average flux.

Observations of PG 1553+113 were conducted by VERITAS from May 2010 to June 2012, totalling 95 hours of observation time. Data were collected in wobble mode, with the source positioned at a 0.5-degree offset from the telescope's pointing direction to enable simultaneous background estimation. The zenith angles for these observations ranged from 20 to 30 degrees, with an average of 23 degrees. The combination of low zenith angles and event selection cuts optimized for soft-spectrum sources resulted in an analysis energy threshold (the energy at which the photon rate peaks after cuts) of 180 GeV.

The analysis used a circular signal region centred on the source's nominal position, extending radially outward by 0.14 degrees. After applying quality selection criteria based on weather conditions and instrument stability, and accounting for instrument read-out dead time, a total of 80 hours of live time were retained. These observations yielded an overall detection significance of 53 standard deviations, with the excess consistent with a point-like source at the location of PG 1553+113.

\section{THE PHOTOHADRONIC MODEL}\label{sec3}

The photohadronic model proposes a double jet configuration during the VHE flaring process~\citep{Sahu:2019lwj,Sahu_2019}. This configuration, previously suggested in earlier studies~\citep{10.1111/j.1365-2966.2008.13360.x,10.1111/j.1365-2966.2009.16045.x}, involves a narrower, more compact jet of size $R'_f$ embedded within a broader jet of size $R'_b$, where $R'_f < R'_b$ (quantities in the comoving frame are denoted by primes). The photon density in the inner jet region, $n'_{\gamma,f}$, is significantly higher than in the outer region, $n'_\gamma$ ($n'_{\gamma,f} \gg n'_\gamma$). The model builds on the conventional interpretation of the first two peaks in the SED: the first peak arises from synchrotron radiation emitted by relativistic electrons in the jet, while the second peak results from the inverse Compton scattering process. 

The inner jet is assumed to move at a slightly higher velocity, characterized by a bulk Lorentz factor $\Gamma_{\rm in}$, compared to the outer jet with $\Gamma_{\rm ext}$. For simplicity, it is assumed that $\Gamma_{\rm in} \simeq \Gamma_{\rm ext} \equiv \Gamma$, and both share a common Doppler factor $\mathcal{D}$~\citep{Ghisellini:1998it,Krawczynski:2003fq}. In the case of HBLs, $\Gamma \simeq \mathcal{D}$.

In this scenario, protons are accelerated to extremely high energies within the inner jet region, following a power-law energy distribution, $dN_p/dE_p \propto E^{-\alpha}_p$~\citep{1993ApJ...416..458D}, where $E_p$ is the proton energy and $\alpha \geq 2$ is the proton spectral index. The value of $\alpha$ varies depending on the type of shock involved, such as non-relativistic, highly relativistic, or oblique relativistic shocks~\citep{2005PhRvL..94k1102K,2012ApJ...745...63S}. These high-energy protons interact with SSC background photons in the inner jet, producing $\Delta$-resonances via the process $p + \gamma \rightarrow \Delta^{+}$. The $\Delta$-resonance decays into neutral and charged pions with different probabilities. While direct single-pion and multi-pion production processes also contribute, they are less efficient in the energy range considered here~\citep{1999PASA...16..160M,2018MNRAS.481..666O}, and their contributions are neglected in this work. The neutral pions decay into gamma-rays ($\pi^0\to\gamma\gamma$), while the charged pions decay into neutrinos ($\pi^+\to e^+\nu_e\nu_\mu\bar{\nu}_\mu$)~\citep{PhysRevD.85.043012}. In the photohadronic model, the gamma-rays from $\pi^0$ decay are blue-shifted to VHE energies and detected on Earth, while the positrons from $\pi^+$ decay emit synchrotron radiation.

The observed VHE gamma-ray energy $E_\gamma$ and the seed photon energy $\epsilon_\gamma$ are related by the condition~\citep{Sahu:2019lwj,Sahu_2019}
\beq
E_{\gamma}\epsilon_{\gamma} \simeq 0.032 \frac{{\cal D}^2}{(1+z)^2}\,\mathrm{GeV}^2.
\label{KinCon}
\eeq
In this process, the VHE photon carries roughly 10 per cent of the proton energy, such that $E_p = 10E_\gamma$. Due to the inaccessibility of the inner jet region and the absence of direct methods to measure the photon density within it, a scaling relationship is assumed between the photon densities in the inner and outer jet regions. This relationship can be expressed as~\citep{Sahu:2015tua}
\beq
\frac{n'_{\gamma,f}(\epsilon_{\gamma,1})}{n'_{\gamma,f}(\epsilon_{\gamma,2})} \simeq \frac{n'_{\gamma}(\epsilon_{\gamma,1})}{n'_{\gamma}(\epsilon_{\gamma,2})},
\label{eq:scalingI}
\eeq
where the left-hand side represents the unknown photon density ratio in the inner jet region, and the right-hand side corresponds to the known ratio in the outer jet region. This equation allows the unknown photon density in the inner region to be expressed in terms of the measurable photon density in the outer region.

The intrinsic gamma-ray flux resulting from the decay of $\pi^0$ particles is derived as
\beq
F_{\rm int}(E_{\gamma}) \equiv E^2_{\gamma} \frac{dN(E_\gamma)}{dE_\gamma} 
\propto  E^2_p \frac{dN(E_p)}{dE_p} n'_{\gamma,f}.
\eeq
By incorporating the seed photon density and the proton flux, the intrinsic flux can be expressed in the form
\beq
F_{\rm int}(E_{\gamma}) = F_0 \left ( \frac{E_\gamma}{\rm TeV} \right )^{-\delta+3},
\label{eq:fluxintrinsic}
\eeq
where $F_0$ is a normalization factor. Substituting this into Eq. (\ref{eq:flux}), the observed flux is given by
\beq
F_{\gamma}(E_{\gamma}) = F_0 \left ( \frac{E_\gamma}{\rm TeV} \right )^{-\delta+3}
e^{-\tau_{\gamma\gamma}(E_{\gamma},z)},
\label{fluxrelat}
\eeq
where $\tau_{\gamma\gamma}(E_{\gamma},z)$ accounts for the absorption of gamma-rays due to interactions with the EBL. The spectral index $\delta$ is defined as $\delta = \alpha + \beta$, where $\alpha$ is the proton spectral index and $\beta$ represents the power-law behaviour of the seed photon spectrum in the low-energy tail of the SSC spectrum. The normalization factor $F_0$ is determined from observational data, while $\delta$ serves as the sole free parameter in the photohadronic model~\citep{Sahu_2019}. As evident from Eq. (\ref{fluxrelat}), any curvature in the spectrum can be attributed to the exponential absorption term~\citep{10.1093/mnras/stz943}.

It is worth emphasizing that the photohadronic process is most effective for gamma-ray energies $E_\gamma \gtrsim 100$ GeV. At lower energies, leptonic processes such as electron synchrotron radiation and the SSC mechanism dominate the multiwavelength SED.

\section{ANALYSIS AND RESULTS}\label{sec4}

We analyse all the seventeen observations discussed in Sec.~\ref{sample} in the context of the photohadronic model to constrain the redshift of the HBL PG 1553+113. To account for the EBL correction to the VHE gamma-ray spectra, we use four well known and widely used EBL models. Here, these models are referred to as Saldana~\citep{2021MNRAS.507.5144S}, Franceschini~\citep{Franceschini:2008tp}, Domínguez~\citep{Dominguez:2010bv}, and Gilmore~\citep{10.1111/j.1365-2966.2012.20841.x}.

Following the classification scheme proposed by~\cite{Sahu_2019}, the VHE emission states of a HBL are defined based on the value of the spectral index $\delta$. A very high emission state corresponds to $2.5 \leq \delta \leq 2.6$, a high emission state to $2.6 < \delta < 3.0$, and a low emission state to $\delta = 3.0$. Since PG 1553+113 is a HBL, $\delta$ is constrained to the range $2.5 \leq \delta \leq 3.0$. Given that the duration of observations varies across datasets, each observation is expected to represent a distinct emission state, necessitating independent values of $\delta$. Therefore, $\delta$ is treated as a constrained but adjustable parameter during the fitting process.

To analyse the VHE spectra and determine the best fits, each EBL model is integrated into the photohadronic framework. A global $\chi^2$ fit is performed on all data points by simultaneously varying the redshift $z$ (common to all observations), the normalization constants $F_0$, and the spectral parameter $\delta$ for each of the seventeen independent observations. This global fitting procedure is repeated for all four EBL models to identify the best-fitting values of $z$, $F_0$, and $\delta$. Confidence level intervals for the statistical errors of the redshift at 68, 90, and 95 per cent of coverage probability are then calculated for each EBL model.

With 58 degrees of freedom (93 experimental data points and 35 free parameters), the minimum $\chi^2$ values obtained for the global fit are 58.82, 65.50, 68.50, and 75.67 for the Saldana, Gilmore, Franceschini, and Domínguez EBL models, respectively. The best-fitting values for $F_0$ and $\delta$ for each observation, along with the corresponding EBL models, are shown in Table~\ref{table2}. Notably, the $F_0$ values exhibit minor variations across EBL models for a given observation, indicating similarities among the models. The best-fitting redshift and its CL intervals at 68, 90, and 95 per cent are presented in Table~\ref{table3}. Each EBL model predicts slightly different CL intervals for $z$, reflecting variations in the models' predictions.

The fits to the seventeen observed VHE spectra of PG 1553+113 are illustrated in Fig.~\ref{fig:Fig1}, using the best-fitting parameters from the Saldana EBL model (detailed in Tables~\ref{table2} and \ref{table3}). Fits based on the Domínguez, Franceschini, and Gilmore EBL models are not shown, as they are visually indistinguishable from those of the Saldana model.

For comparison, all four EBL models are applied to fit the spectrum of H13, as shown in Fig.~\ref{fig:Fig2}. While all models provide excellent fits, a slight discrepancy is observed for $E_{\gamma} \gtrsim 0.3$~TeV. Additionally, Table~\ref{table4} summarizes the redshift constraints for PG 1553+113 estimated by various authors, including those derived in this work.

\section{SUMMARY AND DISCUSSION}\label{sec5}

PG 1553+113 is a luminous, highly variable gamma-ray blazar observed across the electromagnetic spectrum, from radio to TeV energies. Its extreme energy output and relativistic jet make it a pivotal target for probing particle acceleration mechanisms and jet dynamics in AGNs. As one of the brightest TeV emitters, its gamma-ray photons undergo interactions with the EBL during propagation to Earth. These interactions provide critical constraints on EBL models, which are indispensable for advancing the understanding of galaxy evolution and the cosmic infrared background. While PG 1553+113 lacks a spectroscopically confirmed redshift \(z\), estimating this parameter remains vital: \(z\) directly determines the source’s intrinsic luminosity and distance, enabling precise modelling of EBL photon absorption attenuating its gamma-ray emission. Such modelling is essential for testing EBL density predictions.

In this work, the photohadronic model is employed in conjunction with four EBL models—Saldana, Franceschini, Domínguez, and Gilmore—to derive constraints on the redshift of the HBL PG 1553+113. The analysis incorporates seventeen independent VHE spectra collected by the HESS, MAGIC, and VERITAS telescopes between 2005 and 2015. The photohadronic model, which has been successfully applied in previous studies to constrain the redshifts of HBLs with unknown $z$ values~\citep{Sahu:2019lwj,Sahu_2019}, is used here to interpret the VHE spectra of PG 1553+113.

A global $\chi^2$ fitting procedure is performed, simultaneously varying the redshift $z$, the normalization constants $F_0$, and the spectral parameter $\delta$ for each of the seventeen observations, to identify the best fit to the observed VHE spectra for a given EBL model. This process is repeated for the remaining three EBL models to determine the optimal values of $z$, $F_0$, and $\delta$ for each case. Using these best-fitting parameters, CL intervals for the redshift at 68, 90, and 95 per cent are computed for each EBL model.

Table~\ref{table4} presents, in chronological order, the redshift constraints for PG 1553+113 reported in previous studies, alongside those derived in this work. The 95 per cent CL intervals obtained here exhibit substantial overlap with earlier redshift estimates, demonstrating consistency with prior results. Notably, narrower intervals are achieved at high confidence levels in certain cases, attributable to the robust statistical power of the comprehensive dataset utilized in the fitting process, see Fig.~\ref{fig:Fig3}. These outcomes underscore both the reliability of the photohadronic model and the accuracy of the EBL corrections employed in this analysis. The findings confirm that the photohadronic framework accurately models all observed spectra. Among the tested EBL models, the lowest total $\chi^2$ value (58.82 for 58 degrees of freedom) was achieved with the~\cite{2021MNRAS.507.5144S} model, while the~\cite{Dominguez:2010bv} model yielded the highest value ($\chi^2 = 75.67$). The 95 per cent CL intervals for the redshift’s statistical uncertainty are $0.500 < z < 0.537$ (Saldana-Lopez) and $0.491 < z < 0.527$ (Domínguez ).

The photohadronic model, combined with EBL corrections, provides a reliable method for estimating the redshift of PG 1553+113. The results highlight the effectiveness of the photohadronic model in interpreting the VHE gamma-ray spectra and its potential application to other HBLs with unknown redshifts.


\section*{Acknowledgements}
We thank the reviewer for her/his constructive remarks which helped us to improve the manuscript substantially. R. de J. P-A and G. S-C thank SECIHTI (México) for its partial support. Partial support from CSU-Long Beach is gratefully acknowledged. 

\section*{Data Availability}
No new data were generated or analysed in support of this research.


\begin{table*}
\centering
\caption{
Details of the seventeen PG 1553+113 observations used in this study are provided. The first and second columns list the observation name and the instrument used, respectively. The third, fourth, fifth, and sixth columns include the start date, end date, period span (in MJD), and live observation duration (in hours), respectively. The seventh column references the work where the data for each observation period is originally reported.
}
\begin{tabular}{lllllcl}
\hline
Observation	& Instrument & Start date & End date & Period &	Live duration & Reference \\
\hline
H05	&	HESS	&	2005 April	&	2005 August	&	53492-53614	&	7.6	& \cite{2008AA477481A} \\
H06	&	HESS	&	2006 April	&	2006 July	&	53849-53943	&	17.2 & \cite{2008AA477481A} \\
H12	&	HESS	&	2012 April 26	&	2012 April 27	&	56043-56044	&	3.2	& \cite{2015ApJ80265A} \\
H13	&	HESS	&	2013 May 29	&	2013 August 9	&	56441–56513	&	16.8	&	\cite{2017AA600A89H} \\
M0506 &	MAGIC	&	2005 April	&	2006 April	&	--	&	18.8	& \cite{Albert_2007} \\
M06	&	MAGIC	&	2006 July 14	&	2006 July 27	&	53930-53943	& 8.5	& \cite{2009AA493467A} \\
M07	&	MAGIC	&	2007 March 23	&	2007 April 24	&	54182-54214	&	11.5	&	\cite{2012ApJ74846A} \\
M08	&	MAGIC	&	2008 March 17	&	2008 May 7	&	54542-54593	&	8.7	&	\cite{2012ApJ74846A} \\
M08II &	MAGIC	&	2008 March 16	&	2008 April 30	&	54541-54586	&	7.18	&	\cite{2010AA515A76A} \\
M09	&	MAGIC	&	2009 April 16	&	2009 June 15	&	54937-54997	&	8.5	&	\cite{2012ApJ74846A}	\\
M12	&	MAGIC	&	2012 April 20	&	2012 April 26	&	56037-56043	&	1.1	&	\cite{2015MNRAS.450.4399A}	\\
M12II &	MAGIC	&	2012 February 28	&	2012 March 4	&	55985-55990	&	2.4	&	\cite{10.1093/mnras/stz943}	\\
M12III &	MAGIC	&	2012 March 13	&	2012 May 2	&	55999-56049	&	24.84	&	\cite{10.1093/mnras/stz943}	\\
M13	&	MAGIC	&	2013 April 7	&	2013 June 12	&	56389-56455	&	11.42 &	\cite{10.1093/mnras/stz943}	\\
M14	&	MAGIC	&	2014 March 11	&	2014 March 25	&	56727-56741	&	1.93 &	\cite{10.1093/mnras/stz943}	\\
M15	&	MAGIC	&	2015 January 25	&	2015 August 7	&	57047-57241	&	25.77 &	\cite{10.1093/mnras/stz943}	\\
V1012 &	VERITAS	&	2010 May	&	2012 June	&	--	&	80	&	\cite{2015ApJ7997A}	\\
\hline
\end{tabular}
\label{table1}
\end{table*}

\begin{table*}
\centering
\caption{
The best-fitting values for the normalization constant $F_0$ (expressed in units of $10^{-11}\, \mathrm{erg\, cm^{-2}\, s^{-1}}$) and the spectral index $\delta$ of the photohadronic model, determined from fitting the VHE spectra of PG 1553+113, are displayed in the third, fourth, seventh, and eighth columns. These results correspond to four distinct EBL models. The first and fifth columns identify the observations, while the second and sixth columns specify the EBL models applied, as outlined in the main text.
}
\begin{tabular}{clcclclcc}
\hline
Observation & EBL Model & $F_0$ & $\delta$ & &
Observation & EBL Model & $F_0$ & $\delta$  \\
\hline
H05	&	Domínguez	&	11.14	&	2.50	& &
H06	&	Domínguez	&	7.83	&	2.50	\\
	&	Saldana	&	12.87	&	2.50	& &
	&	Saldana	&	9.05	&	2.50	\\
	&	Franceschini	&	11.08	&	2.50	& &
	&	Franceschini	&	7.78	&	2.50	\\
	&	Gilmore	&	12.15	&	2.50	& &
	&	Gilmore	&	8.54	&	2.50	\\
H12	&	Domínguez	&	37.91	&	2.50	& &
H13	&	Domínguez	&	14.29	&	2.66	\\
	&	Saldana	&	43.41	&	2.50	& &
	&	Saldana	&	18.28	&	2.59	\\
	&	Franceschini	&	37.60	&	2.50	& &
	&	Franceschini	&	14.71	&	2.63	\\
	&	Gilmore	&	41.16	&	2.50	& &
	&	Gilmore	&	16.37	&	2.62	\\
M0506 &	Domínguez	&	3.82	&	3.00	& &
M06	&	Domínguez	&	2.82	&	3.00	\\
	&	Saldana	&	4.39	&	3.00	& &
	&	Saldana	&	3.24	&	3.00	\\
	&	Franceschini	&	3.78	&	3.00	& &
	&	Franceschini	&	2.80	&	3.00	\\
	&	Gilmore	&	4.15	&	3.00	& &
	&	Gilmore	&	3.08	&	3.00	\\
M07	&	Domínguez	&	2.27	&	3.00	& &
M08	&	Domínguez	&	5.22	&	2.96	\\
	&	Saldana	&	2.63	&	3.00	& &
	&	Saldana	&	6.23	&	2.93	\\
	&	Franceschini	&	2.25	&	3.00	& &
	&	Franceschini	&	5.30	&	2.94	\\
	&	Gilmore	&	2.48	&	3.00	& &
	&	Gilmore	&	5.78	&	2.95	\\
M08II	&	Domínguez	&	4.80	&	2.97	& &
M09	&	Domínguez	&	5.23	&	2.50	\\
	&	Saldana	&	5.88	&	2.93	& &
	&	Saldana	&	6.03	&	2.50	\\
	&	Franceschini	&	4.88	&	2.95	& &
	&	Franceschini	&	5.18	&	2.50	\\
	&	Gilmore	&	5.39	&	2.95	& &
	&	Gilmore	&	5.70	&	2.50	\\
M12	&	Domínguez	&	10.00	&	2.94	& &
M12II	&	Domínguez	&	5.71	&	3.00	\\
	&	Saldana	&	12.39	&	2.88	& &
	&	Saldana	&	6.41	&	3.00	\\
	&	Franceschini	&	10.06	&	2.92	& &
	&	Franceschini	&	5.62	&	3.00	\\
	&	Gilmore	&	11.21	&	2.91	& &
	&	Gilmore	&	6.14	&	3.00	\\
M12III	&	Domínguez	&	8.77	&	2.88	& &
M13	&	Domínguez	&	4.50	&	3.00	\\
	&	Saldana	&	10.71	&	2.84	& &
	&	Saldana	&	5.27	&	2.97	\\
	&	Franceschini	&	8.89	&	2.87	& &
	&	Franceschini	&	4.42	&	3.00	\\
	&	Gilmore	&	9.83	&	2.86	& &
	&	Gilmore	&	4.82	&	3.00	\\
M14	&	Domínguez	&	5.59	&	3.00	& &
M15	&	Domínguez	&	4.70	&	3.00	\\
	&	Saldana	&	6.19	&	3.00	& &
	&	Saldana	&	5.26	&	3.00	\\
	&	Franceschini	&	5.49	&	3.00	& &
	&	Franceschini	&	4.62	&	3.00	\\
	&	Gilmore	&	5.98	&	3.00	& &
	&	Gilmore	&	5.05	&	3.00	\\
V1012	&	Domínguez	&	6.16	&	2.83	& &
	&		&		&		\\
	&	Saldana	&	6.99	&	2.84	& &
	&		&		&		\\
	&	Franceschini	&	6.20	&	2.82	& &
	&		&		&		\\
	&	Gilmore	&	6.72	&	2.83	& &
	&		&		&		\\
\hline
\end{tabular}
\label{table2}
\end{table*}

\begin{table*}
\centering
\caption{
The redshift values derived from the photohadronic model, based on a global $\chi^2$ fit to the VHE spectra of PG 1553+113, are summarized. The first column identifies the EBL models used, as discussed in the main text. The second column provides the best-fitting redshift $z$, while the third, fourth, and fifth columns present the CL intervals for the statistical error at 68, 90, and 95 per cent, respectively, for each EBL model.
}
\begin{tabular}{lcccc}
\hline
EBL Model & Redshift & \multicolumn{3}{c}{Redshift CL intervals} \\
 & $z$ & $68\%$ & $90\%$ & $95\%$ \\
\hline
Domínguez	&	0.509	&	(0.500, 0.518)	&	(0.494, 0.524)	&	(0.491, 0.527)	\\
Saldana	&	0.518	&	(0.509, 0.528)	&	(0.503, 0.534)	&	(0.500, 0.537)	\\
Franceschini	&	0.515	&	(0.506, 0.524)	&	(0.500, 0.530)	&	(0.497, 0.533)	\\
Gilmore	&	0.557	&	(0.547, 0.566)	&	(0.541, 0.573)	&	(0.538, 0.576)	\\
\hline
\end{tabular}
\label{table3}
\end{table*}

\begin{table*}
\centering
\caption{
A compilation of redshift estimates for PG 1553+113 from various studies is presented. The first column lists the redshift measurements, including interval limits or central values with associated uncertainties. The second column describes the methodologies employed to determine the redshift, while the third column cites the relevant references.
}
\begin{tabular}{cll}
\hline
Redshift & Method & Reference \\
\hline
$0.491 < z < 0.527$ at 95 &  Global fit of the photohadronic model to independent & This work  \\ per cent CL	&  observations (EBL Domínguez) &  \\
$0.500 < z < 0.537$ at 95 &  Global fit of the photohadronic model to independent & This work  \\ per cent CL	&  observations (EBL Saldana) &  \\
$0.408 < z < 0.436$	&	Edge of the Ly\ $\alpha$ forest observed in UV spectra	& \cite{2022MNRAS.509.4330D} \\
$z$ = 0.38 ± 0.10	& VHE spectral index (lowest-flux state) -- redshift correlation &	\cite{10.1093/mnras/stab3173} \\
$z$ = 0.33 ± 0.08	& VHE spectral index (average) -- redshift correlation &	\cite{10.1093/mnras/stab3173} \\
$0.22<z<0.48$	&	Leptonic model fit to multiple SEDs	&	\cite{2018MNRAS.473.3755Q}	\\
$0.4339\leq z$	& Analysis of intervening OVII absorbers along the line of sight of the object &	\cite{nicastro2018confirmingdetectionwhimsystems}	\\
$0.3551\leq z$	& Analysis of intervening OVII absorbers along the line of sight of the object &	\cite{nicastro2018confirmingdetectionwhimsystems}	\\
$z = 0.49 \pm 0.04$	&	Comparison of the hardness of different SEDs &	\cite{2015ApJ80265A}	\\
$z\leq 0.62$	& Energy spectrum and EBL evolution	&	\cite{2015ApJ7997A}	\\
$0.3<z$	& X-Shooter spectroscopy	&	\cite{refId05}	\\
$z=0.75^{+0.04}_{-0.05}$	&	Stepwise spectrum deabsortion and minimum $\chi^2$ fit	&	\cite{2010ApJ7081310A}	\\
$0.43 < z \lesssim 0.58$	&	Detection (non-detection) of Ly\ $\alpha$ (Ly\ $\beta$) absorbers & \cite{2010ApJ720976D}	\\
$z<0.78$	&	Intrinsic and absorbed spectra comparison	&	\cite{2010PASJ62L23Y}	\\
$z<0.67$	&	Constriction of the hardness of the spectrum	&	\cite{2009arXiv0907.0157P}	\\
$z<0.69$	&	Parameter variation deabsortion	& \cite{2008AA477481A} \\
$z<0.69$	& Intrinsic photonic index limit &	\cite{2007ApJ655L13M}	\\
$0.3<z<0.4$	&	Constraint on the apparent magnitude of the host galaxy	&	\cite{2007AA473L17T}	\\
$z<0.74$	&	Parameter variation deabsortion	&	\cite{Albert_2007}	\\
$z<0.74$	&	Spectrum deabsortion	&	\cite{2006AA448L19A}	\\
\hline
\end{tabular}
\label{table4}
\end{table*}

\begin{figure*}
\centering
\includegraphics[width=7in]{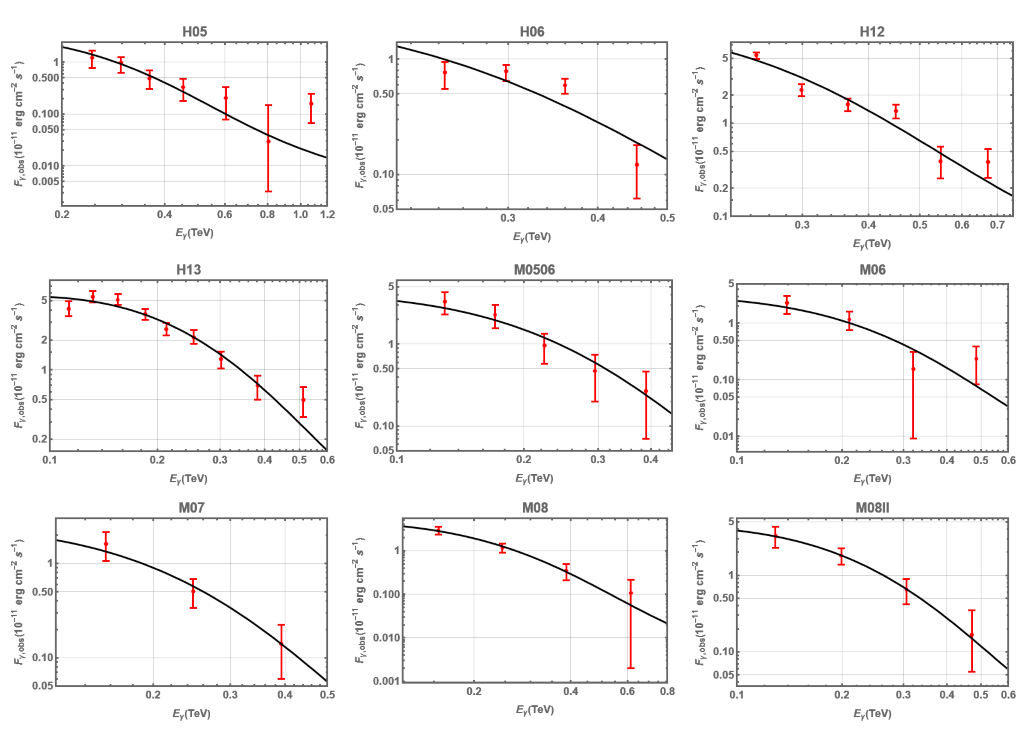}
\caption{
The photohadronic model, combined with the Saldana EBL model, is applied to fit the seventeen observed VHE gamma-ray spectra of PG 1553+113. The corresponding redshift is fixed at $z=0.518$, while the normalization constants $F_0$ and spectral parameters $\delta$ are provided in Table~\ref{table2}.
}
\label{fig:Fig1}
\end{figure*}

\begin{figure*}
\ContinuedFloat
\centering
\includegraphics[width=7in]{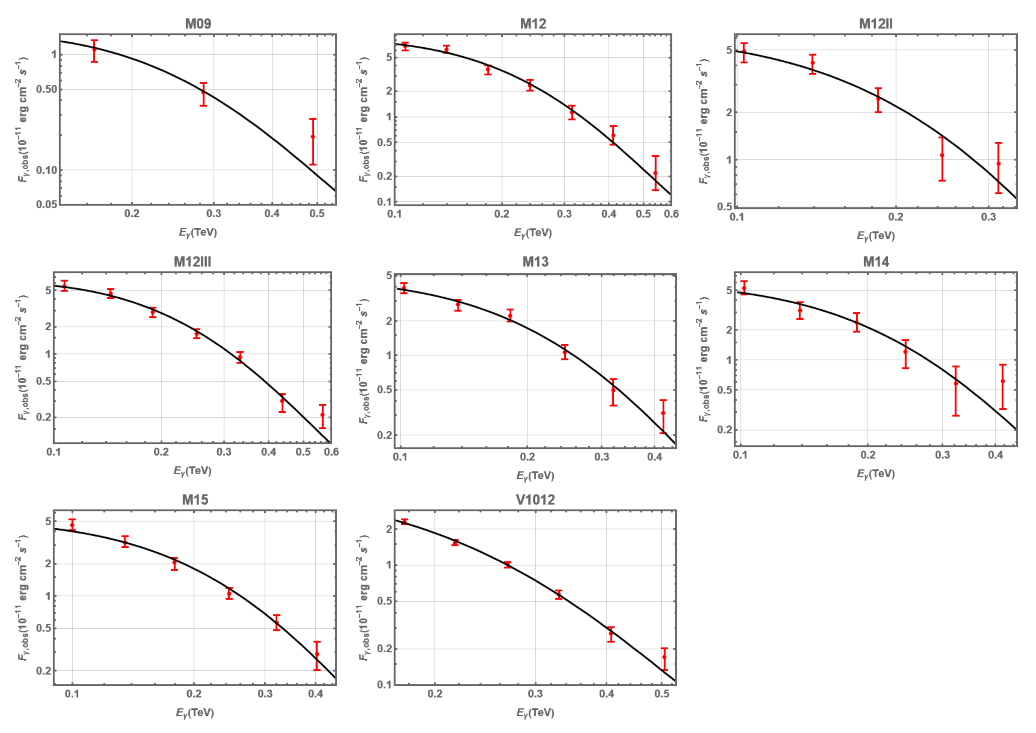}
\caption{
(Continued) The photohadronic model, combined with the Saldana EBL model, is applied to fit the seventeen observed VHE gamma-ray spectra of PG 1553+113. The corresponding redshift is fixed at $z=0.518$, while the normalization constants $F_0$ and spectral parameters $\delta$ are provided in Table~\ref{table2}.
}
\end{figure*}

\begin{figure*}
\centering
\includegraphics[width=6in]{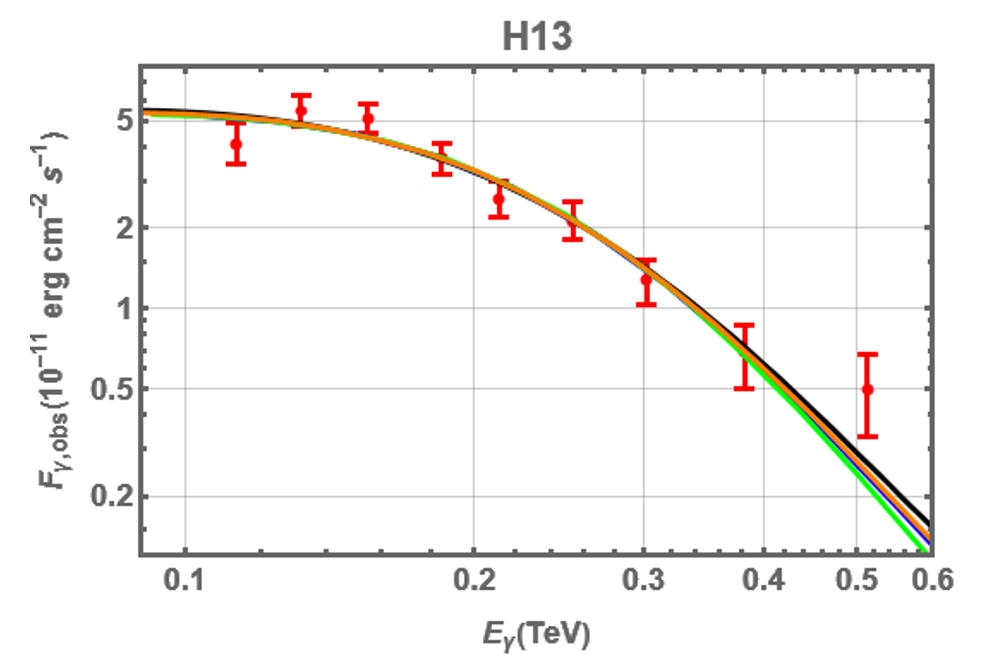}
\caption{
Fits of the photohadronic model to the observed VHE spectrum H13 of PG 1553+113 are presented to compare the performance of various EBL models: Saldana (black), Franceschini (blue), Domínguez (green), and Gilmore (orange). The associated values for the spectral parameter $\delta$, normalization constants $F_0$, and redshift $z$ can be found in Tables~\ref{table2} and~\ref{table3}.
}
\label{fig:Fig2}
\end{figure*}

\begin{figure*}
\centering
\includegraphics[width=6in]{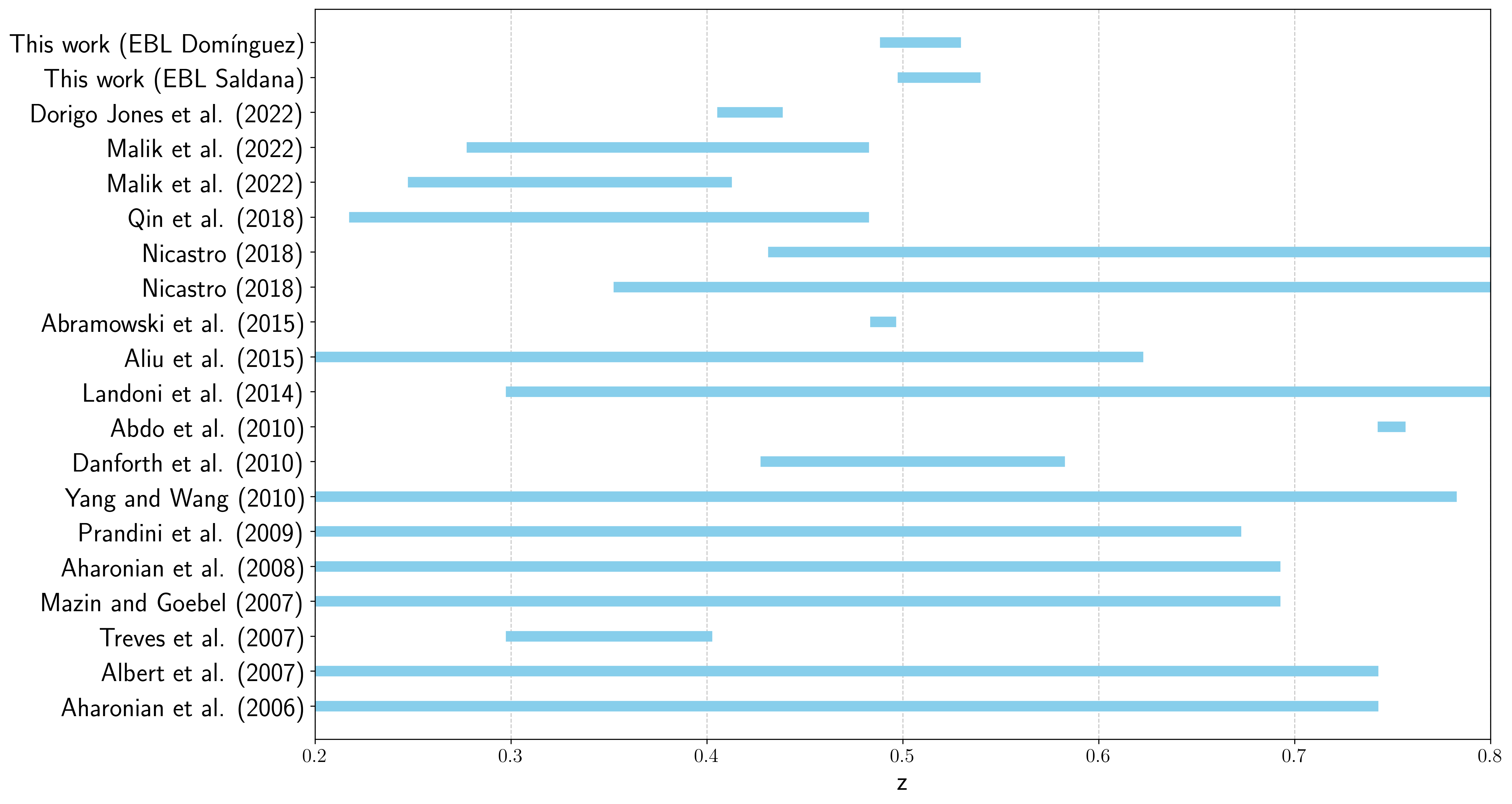}
\caption{
Graphical comparison of historical redshift estimates for PG 1553+113, which complements the quantitative constraints presented in Table~\ref{table4}. The figure provides a consolidated overview of the redshift ranges reported in prior studies alongside those derived in this work.
}
\label{fig:Fig3}
\end{figure*}



\bibliographystyle{mnras}
\bibliography{ref}


\bsp	
\label{lastpage}
\end{document}